# Evolving Delta-oriented Software Product Line Architectures

Arne Haber[1], Holger Rendel[1],
Bernhard Rumpe[1], and Ina Schaefer[2]

[1] Software Engineering, RWTH Aachen University, Germany
[2] Institute for Software Systems Engineering, TU Braunschweig, Germany

**Abstract.** Diversity is prevalent in modern software systems. Several system variants exist at the same time in order to adapt to changing user requirements. Additionally, software systems evolve over time in order to adjust to unanticipated changes in their application environment. In modern software development, software architecture modeling is an important means to deal with system complexity by architectural decomposition. This leads to the need of architectural description languages that can represent spatial and temporal variability. In this paper, we present delta modeling of software architectures as a uniform modeling formalism for architectural variability in space and in time. In order to avoid degeneration of the product line model under system evolution, we present refactoring techniques to maintain and improve the quality of the variability model. Using a running example from the automotive domain, we evaluate our approach by carrying out a case study that compares delta modeling with annotative variability modeling.

## 1 Introduction

Modern software systems simultaneously exist in many different variants in order to adapt to changing user requirements or application contexts. Software product line engineering [32] aims at developing a family of systems by managed reuse in order to decrease time to market and to improve quality. In addition to this variability in space, software systems are extremely long-lived and have to evolve over time in order to maintain, improve or update their functionality. This unanticipated variability in time [26] changes the system design, structure, and behavior in an unexpected manner, e.g., for adapting it to new customer requirements or technological conditions. Evolution of software systems needs to be managed, and gets particularly difficult if a family of systems is considered.

The design of the software architecture plays an essential role in software development [27, 25]. The architecture allows decomposing a complex system into smaller hierarchically structured components. These can be developed independently. The change frequency of architectural descriptions is lower than the changes on the implementation level, where often bugs etc. need to be fixed. However, changes in the architecture have a wide range impact on the overall system such that architectural changes have to be planned, modeled and analyzed to ensure that the system quality is maintained despite of the changes. This is particularly complex for software product line architectures.

Most current ADLs [25] do not support the explicit representation of architectural change. The predominantly used approaches for architectural variability modeling use annotations to assign model elements to different variants. These *annotative variability modeling* approaches mostly use a so called 150%-percent model of the system architecture incorporating all possible variability in which specific elements are annotated to belong to specific product variants. The monolithic 150%-percent architecture description gets easily very complex for large product families and is hard to manage in case of evolutionary changes. Introducing a new variant will most likely require changes of the whole model, as modular development and implementation of variable parts is not possible. To counter this problem, ADLs should support variability modeling by representing changes to the architecture in space and in time as explicit first-class entities. The variability description in the ADL should be modular to facilitate tracing changes to particular functions, components, or features. Furthermore, the description should be readable, easy to comprehend, to evolve, and to maintain.

In this paper, we present $\Delta$-MontiArc, an ADL with native support for architectural variability modeling in space and in time that allows defining variants of interactive distributed and Cyber-Physical systems in a modular manner. $\Delta$-MontiArc is based on the concept of delta modeling software product lines [6]. A product line of architectures is described by a core architecture and a set of architectural deltas that encapsulate changes to the core architecture. In order to obtain a particular product variant, a set of suitable deltas defined in a product configuration is applied to the core. As variable parts of a model, e.g. functionality for new product variants, are encapsulated in deltas, this approach overcomes the aforementioned problems of annotative variability modeling. As complexity of models is decreased and modular modeling of variability is possible, delta models are easier to comprehend and to evolve. In previous work [16, 14], $\Delta$-MontiArc was used to represent spatial variability only. In this paper, we extend it to capture temporal variability with the same linguistic means. If new products should be included in a product family, new deltas can easily be added to a delta model to generate new variants. If a product variant is no longer supported, its product configuration and redundant deltas may be removed. Modifications to certain product functionalities, e.g., for bug fixing, can be realized by replacing a particular delta by another version. In order to avoid degeneration of a delta model after some evolution steps, it can be refactored to improve its structure without changing the generated products. The evolution of architectures as considered in this paper reflects the evolution of the features contained in a software product line. However, the presented approach solely works on the level of the product line artifacts modeling solution space variability [8], in contrast to problem space variability that is typically captured with product features on the requirements level.

In order to evaluate $\Delta$-MontiArc, we carried out a case study to gain experience in spatial and temporal evolution of delta oriented product lines. This case study has been also modeled using a common annotative variability modeling approach to compare it with our approach. The case study describes a braking controller system which exists in variants for cars and motorcycles and allows the inclusion of several assistance system, like an anti-lock braking system or an electronic stability control. By considering several evolution and refactoring scenarios, we demonstrate that delta modeling is particularly well suited for representing architectural variability and architectural evolution.

The paper is structured as follows. Sect. 2 introduces $\Delta$-MontiArc for representing spatial architectural variability. Sect. 3 demonstrates how $\Delta$-MontiArc captures temporal architectural variability. Sect. 4 shows three refactoring strategies for delta models. Sect. 5 contains a qualitative and quantitative comparison of $\Delta$-MontiArc and an annotative variability modeling approach based on the preformed case study. Related work is discussed in Sect 6. Sect. 7 concludes the paper and outlines future work.

## 2 Spatial Variability

*Delta modeling* [6, 36] is a language-independent approach for modular modeling of variability in the solution space [8] and can be applied to different modeling and programming languages like, e.g., class diagrams [35] or Java [34, 36]. In [16, 14], the concept of delta modeling is applied to software architectures in order to obtain an ADL with native support for architectural variability in space. A $\Delta$-MontiArc product line is specified by a designated core architecture that represents the architecture of a valid product variant, and a set of deltas that add, remove, or modify architectural elements to derive further product variants. An architectural variant is definied by a *variant configuration* that contains a set of application-specific deltas that are used in order to generate the variant. Therefore, the operations of these deltas are stepwise applied in a calculated order to transform the core to the architectural variant. After a variant is generated, its correctness is checked using mechanisms of the base language. To with an *application order constraint* (AOC) that determines which deltas must or must not be applied before. If, for example, a delta $A$ modifies a model element, that has been introduced by another delta $B$ and is not part of the core, the AOC of $A$ has to claim that it must be applied after $B$. Hence, the application order of the deltas contained in a variant configuration is calculated by interpreting the attached AOCs. If more then one application order is valid for a product variant, all application orders are expected to generate the same product, not regarding the order of the model elements in the resulting variant. This is the case, if for example two or more deltas of a configuration do not have an attached AOC and their position in the application order may be arbitrary switched without influencing the generated product. However, it is yet unchecked, if several valid application orders really result in the sematically same product. Therefore the correctness of AOCs is assumed. According to [34] it is also possible to define product lines based on more than one valid core architecture. Then, however, the core model that is to be modified must be explicitly referenced in product configurations.

```
component BrakingSystem {
  autoconnect port;

  port
    in BrakeCommand brake,
    out BrakePressure wheelpressure1,
    out BrakePressure wheelpressure2,
    out BrakePressure wheelpressure3,
    out BrakePressure wheelpressure4;

  component PressureCalculator brakefunction;
}
```

**Listing 1.** Core architecture of BrakingSystem.

*Δ*-MontiArc is based on the textual architectural description language (*ADL*) MontiArc [18] that allows modeling and simulation of interactive distributed and Cyber-Physical systems. Therefore it provides modeling elements to describe component type definitions that contain an interface description, an internal structure given by subcomponents, and the communication between subcomponents and the components interface. An example of a MontiArc architecture is given in Lst. 1. It depicts the definition of component type `BrakingSystem` that calculates the brake pressure for all four wheels of a car. MontiArc components communicate with their environment using their interface. The interface definition of component `BrakingSystem` is given by an incoming port `brake` with type `BrakeCommand` (l. 5) and four outgoing ports to emit the calculated brake pressure for each wheel (ll. 6–9). The `BrakingSystem` component contains a subcomponent `brakefunction` that is an instance of component type `PressureCalculator` (l. 11). The connections between the outer ports and the interface of the `brakefunction` subcomponent are created automatically using MontiArc's `autoconnect` statement (l. 2). Parametrized with keyword `port`, it automatically creates connections between all yet unconnected type-compatible ports with the same name.

*Δ*-MontiArc extends the MontiArc ADL with the concepts of delta modeling. Therefore it defines a language that allows modifications of component type definitions by adding or removing model elements like ports, subcomponents, or connections. Lst. 2 shows delta `DTractionControl` specified in *Δ*-MontiArc that adds a traction control functionality to component `BrakingSystem` by modifying the `BrakingSystem` component (c.f. ll. 3 ff). The delta adds an additional port `accel` to receive accelerate commands (l. 4). This port is implicitly connected to the added subcomponent `stabilizer` (l. 5). The aforementioned connections between the interface of component `BrakingSystem` and its sub-

```
1  delta DTractionControl after
2      DAntiLockBrakingSystem  && !DTwoWheel {
3    modify component BrakingSystem {
4      add port in AccelerateCommand accel;
5      add component TC stabilizer;
6
7      connect brakefunction.wheelpressure1 ->
8        stabilizer.fromabs1;
9      connect brakefunction.wheelpressure2 ->
10       stabilizer.fromabs2;
11     connect brakefunction.wheelpressure3 ->
12       stabilizer.fromabs3;
13     connect brakefunction.wheelpressure4 ->
14       stabilizer.fromabs4;
15   }
16 }
```

**Listing 2.** Delta adding traction control.

component `brakefunction` are now explicitly redirected to the newly added subcomponent that itself is implicitly connected to the outer interface (c.f. ll. 7–14). In the example, the AOC given by keyword `after` in ll. 1 f defines that delta `DTractionControl` has to be applied after delta `DAntiLockBrakingSystem` (see Lst. 3) and not before delta `DTwoWheel`. To efficiently check the applicability of deltas and the consistency of the application order constraints during product generation, a family-based analysis of delta-oriented product lines is presented in [15].

Concrete product variants are defined in $\Delta$-MontiArc by a product configuration that specifies which deltas have to be applied to the core architecture to generate a product variant. Lst. 4 shows product configuration `CarWithTC` for a braking system variant that contains an anti-lock braking system (added by delta `DAntiLockBrakingSystem`, see Lst. 3) and a traction control (added by delta `DTractionControl`, see Lst. 2) beside the basic brake functionality introduced by the core architecture which is depicted in Lst. 1.

```
1  delta DAntiLockBrakingSystem {
2    modify component BrakingSystem {
3      add port in WheelSensor wheelspeed1,
4               in WheelSensor wheelspeed2,
5               in WheelSensor wheelspeed3,
6               in WheelSensor wheelspeed4;
7      replace component brakefunction
8        with component ABS brakefunction;
9    }
10 }
```

**Listing 3.** Delta adding anti-lock braking system.

```
1 deltaconfig CarWithTC {
2   DAntiLockBrakingSystem,
3   DTractionControl
4 }
```

**Listing 4.** Product configuration `CarWithTC`.

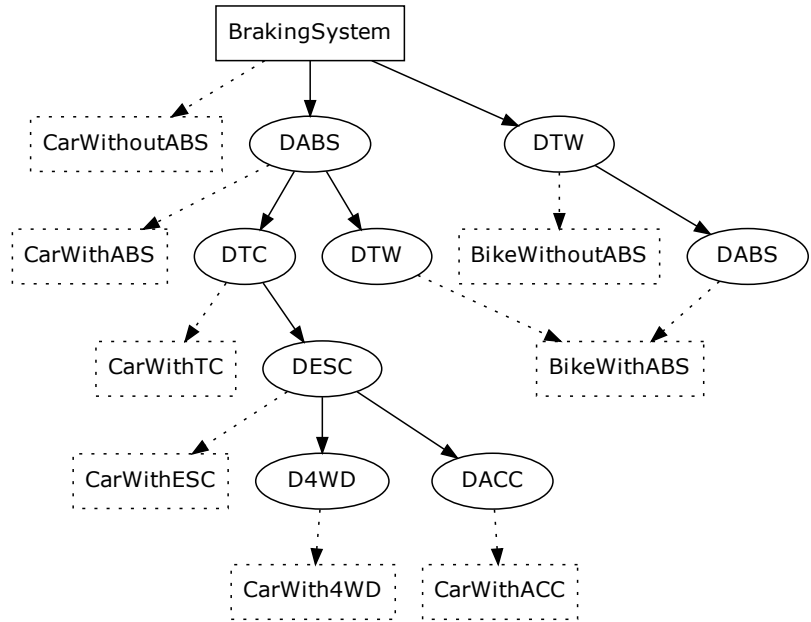

**Fig. 5.** Initial product line structure

As running example throughout this paper, we consider a product line of braking system controllers. In Fig. 5, the delta model of this controller product line is shown. The core architecture `BrakingSystem` (see Lst. 1) is depicted at the very top of the figure. Below, we see all deltas denoted by ellipses. The solid arrows show the possible application orders of the deltas according to the application order constraints in the `after` clauses. The supported product variants are shown in dashed boxes. The product line contains a braking system for cars without an anti-lock braking system (`CarWithoutABS`) as core architecture. By applying the delta `DAntiLockBrakingSystem` (short: DABS), a product variant `CarWithABS` can be obtained. Using the delta `DTwoWheel` (short: DTW), a braking system for motorbikes with only two wheels is generated (`BikeWithoutABS`). Using the delta `DABS`, a braking system for motorbikes with ABS is derived (`BikeWithABS`). For cars, a traction control can be added by delta `DTractionControl` (short: DTC), and afterwards an electronic stability control can be added by delta `DElectronicStabilityControl` (short: DESC). Finally, the architecture can be tailored to work with an adaptive cruise control system by applying delta `DAdaptiveCruiseControl` (short: DACC) or alternatively by a four wheel drive using delta `DFourWheelDrive` (short: D4WD). The initial product line of braking systems realized in $\Delta$-MontiArc supports eight product variants with six deltas.

## 3 Temporal Variability

The difference between temporal and spatial variability is that spatial variability is anticipated and, thus, can be planned ahead while temporal variability is unanticipated and has to be integrated into the product line after its initial design. However, variability in time can be presented by the same means as variability in space using the concepts of delta modeling [6].

The evolution of a product line can be completely classified into three different scenarios: first, new product variants are added; second, product variants are removed; third, existing product variants are modified. In the following, we illustrate how these three evolution scenarios can be represented with $\Delta$-MontiArc without re-engineering the delta models from scratch, but by evolving it via modular and local changes to deltas and product configurations.

*Add Variants.* A delta model in $\Delta$-MontiArc consists of a designated core architecture, a set of architectural deltas and the set of supported product configurations which are selected subsets of the available deltas. When new architectural variants are added, this amounts to adding the respective deltas and product configurations that are required to generate the new product variants which are not yet contained in the delta model.

In our running example, we can add a new product variant to the braking system controller product line that includes support for a reduction gear. This variant is only for driving offroad and, thus, requires that four wheel driving is included in the product as well. To capture this change, a new delta `DReductionGear` (short: DRG) shown in Lst. 6 is added to the delta model. A new configuration `CarWithRG` (see Lst. 7) defines the new product variant.

*Remove Variants.* When product variants are removed, since they are now longer supported or maintained, the respective product configurations can simply be removed from the set of product configurations. If deltas are no longer required for product generation, because all product configurations using them have been removed, also the redundant deltas can be removed. The removal of deltas can require a modification of application order constraints of other remaining deltas. This can only be the case, if the removed delta is mentioned in the `after` clause as a conflicting delta that may not be applied together with this delta. Hence, constraints on removed deltas can be deleted without changing the remaining product variants.

```
1 delta DReductionGear after DFourWheelDrive {
2   modify component BrakingSystem {
3     add component BrakeAmplifier;
4     connect stabilizer.wheelpressure1
5       -> BrakeAmplifier.wheelpressurefromesp1;
6     connect stabilizer.wheelpressure2
7       -> BrakeAmplifier.wheelpressurefromesp2;
8     connect stabilizer.wheelpressure3
9       -> BrakeAmplifier.wheelpressurefromesp3;
10    connect stabilizer.wheelpressure4
11      -> BrakeAmplifier.wheelpressurefromesp4;
12  }
13 }
```

Listing 6. Delta for adding reduction gear.

```
1 deltaconfig CarWithRG {
2   DAntiLockBrakingSystem,
3   DTractionControl,
4   DElectronicStabilityControl,
5   DFourWheelDrive,
6   DReductionGear
7 }
```

Listing 7. Configuration for product variant with reduction gear.

In our running example, assume that the variants `CarWithoutABS` and `CarWithTC` should not be supported anymore, since all cars should now contain either ABS or ESC right away. These configurations can be removed from the product line without changing any delta, since all deltas are still required to generate the remaining variants. Now we assume, that the product portfolio should be consolidated such that only control units for cars are produced and motorbikes are not supported anymore. Hence, the variants `BikeWithoutABS` and `BikeWithABS` are removed and also delta `DTwoWheel` is removed since it is no longer required for generating a product variant. In delta `DTractionControl`, the negated reference to delta `DTwoWheel` is also deleted.

*Modify Variants.* The modification of existing product variants requires to change the implementation of one or more existing deltas. A reason for a modification of an existing delta may, for instance, be a bug fix or an improvement of performance by new component realizations.

In our running example, assume that the existing delta `DAdaptiveCruise-Control` (see Lst. 8) has to be modified by adding a new input port for a rainsensor which is necessary for its correct functioning. The new version of delta `DAdaptiveCruiseControl` is depicted in Lst. 9. As only the implementation inside this delta is changed, the general structure of the product line does not change. From now on, the new delta is used when generating product variants, such that the new corrected functionality of the adaptive cruise control system

is contained in any newly generated product variant. Fig. 10 shows the structure
of the product line after applying all three scenarios. Type safety of all deltas
of a product line may be assured using a family-based analysis depending on
MontiArcs checking facilities as described in [15] or by designing a constraint-
based type system similar to the one presented in [33].

```
delta DAdaptiveCruiseControl after
    DElectronicStabilityControl && !DFourWheelDrive {
  modify component BrakingSystem {
    add port in AccelerateCommand accelfromacc,
            in BrakeCommand brakefromacc;
    add component SignalHandler;
    connect accel -> SignalHandler.accelfromdriver;
    connect brake -> SignalHandler.brakefromdriver;
  }
}
```
**Listing 8.** Original delta for adding adaptive cruise control.

```
delta DAdaptiveCruiseControl after
    DElectronicStabilityControl && !DFourWheelDrive {
  modify component BrakingSystem {
    add port in AccelerateCommand accelfromacc,
            in BrakeCommand brakefromacc,
            in RainIntensity rainsensor;
    add component SignalHandler;
    connect accel -> SignalHandler.accelfromdriver;
    connect brake -> SignalHandler.brakefromdriver;
  }
}
```
**Listing 9.** Modified delta for adaptive cruise control system.

## 4 Refactoring Delta-oriented Product Lines

As we can observe in the previous section, evolving a delta-oriented product
line includes the addition and removal of deltas and the addition and removal
of product configurations. This may lead to a degeneration of the product line
structure, e.g., deltas are factually separated, but always applied together, or
sequences of deltas are always applied to the core without generating individ-
ual products. While this is not a problem for the generated product variants
themselves, it unnecessarily complicates the product line structure and hinders
further evolution and maintenance.

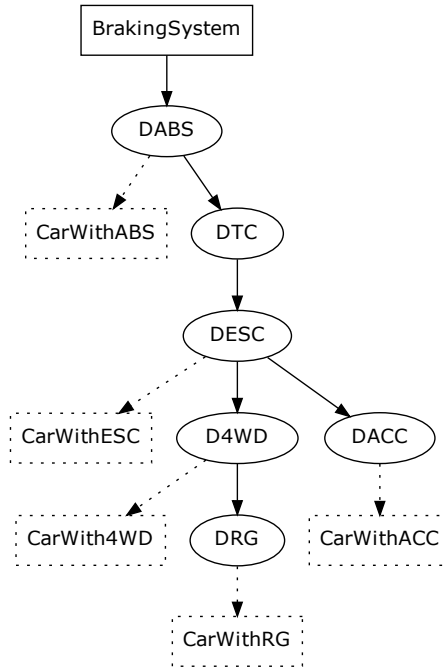

**Fig. 10.** Brake controller product line structure after evolution

Refactoring [11] is a well-known technique on the programming language level to improve the structure of code without changing its semantical meaning. The same idea can also be applied to product lines realized with $\Delta$-MontiArc. *Product line refactorings* aim at reducing the overall complexity of the product line structure and at the same time increasing its comprehensibility. This is achieved by carefully changing the structure of a product line, but preserving the set of products that can be generated. Changes of the structure are accomplished by modifications of

1. the set of available deltas,
2. the content of existing deltas,
3. the application order constraints attached to deltas, and
4. the set of variant configurations.

In this section, we propose exemplary refactoring strategies to maintain the quality of the product line structure after product line evolution. In particular, we consider the *Compose-Deltas-Refactoring* where deltas that are always applied together are merged, the *Merge-With-Core-Refactoring* where deltas are integrated in the core to form a new core and *Merge-With-Core-Refactoring With*

*Inverse Deltas* that extends the possibilities of the former refactoring. This set of refactorings is not complete. Depending on the structure of a product line, more refactoring strategies might be possible.

*Compose Deltas-Refactoring.* The Compose-Deltas-Refactoring merges the content of a sequence of deltas and forms a new delta that contains the combined modifications of the delta sequence.

**Situation:** The precondition for this refactoring is that we have a sequence of deltas that are always applied together and where the intermediate products after applying any prefix of the sequence do not correspond to a supported product variant.

**Mechanics:** The Compose Deltas-Refactoring is carried out as follows:

1. Identify a sequence of deltas $D_1, ..., D_n$ satisfying the above conditions.
2. Construct a new delta $D_n$ containing the modifications of the delta sequence:
   (a) Merge the modification operations of the composed deltas into $D_n$ by putting all delta operations in sequence, starting from $D_1$ and ending with $D_n$. Delta operations targeting the same architectural element can be composed to a single operation. For example, if a component is first added, then removed in a subsequent delta, and finally added again, the three operations can be replaced by a single add operation.
   (b) Compute the new application order constraint $AOC_n$ of $D_n$ which is the union of the application order constraints of the merged deltas $AOC_1, ..., AOC_n$ where references to the deltas $D_1, ..., D_n$ are removed.
3. Adjust all supported product configurations that include the delta sequence $D_1, ..., D_n$ to only include $D_n$.
4. Remove $D_1, ..., D_{n-1}$ from the delta model, since they are no longer used to generate any product variant.

**Effect:** By the Compose-Deltas-Refactoring, product generation is simplified as only one delta instead of a sequence of deltas has to be applied. Additionally, the complexity of the product line decreases since deltas that are no longer required after the refactoring can be removed.

```
delta DElectronicStabilityControl after
    DTractionControl && !DFourWheelDrive {
  modify component BrakingSystem {
    add port in LateralAcceleration lateralaccel;
    replace component stabilizer with component ESC stabilizer;
  }
}
```

**Listing 11.** Delta for adding an electronic stability control system.

**Example:** In our running example, the deltas DTractionControl (set Lst. 2) and DElectronicStabilityControl (see Lst. 11) are always used

```
delta DElectronicStabilityControl after
    DAntiLockBrakingSystem && !DFourWheelDrive{
  modify component BrakingSystem {
    add port in AccelerateCommand accel,
             in LateralAcceleration lateralaccel;
    add component ESC stabilizer;

    connect brakefunction.wheelpressure1 ->
      stabilizer.fromabs1;
    connect brakefunction.wheelpressure2 ->
      stabilizer.fromabs2;
    connect brakefunction.wheelpressure3 ->
      stabilizer.fromabs3;
    connect brakefunction.wheelpressure4 ->
      stabilizer.fromabs4;
  }
}
```

**Listing 12.** Delta composed from DTractionControl and DElectronicStabilityControl.

together and the intermediate product after applying delta DTractionControl is not a supported product variant (see Fig. 10). To simplify the structure, these two deltas may be composed to a single delta which is again called DElectronicStabilityControl and shown in Lst. 12. It contains the delta operations of the two original deltas for adding the ports accel and lateralaccel and the respective connections. For the component stabilizer, there is only one delta operation adding the version of the component introduced by delta DElectronicStabilityControl. Delta DTractionControl adds subcomponent stabilizer to BrakingSystem (l. 5) that is replaced subsequently in the original delta DElectronicStabilityControl by another subcomponent (l. 4). Hence, in the composed delta it suffices to add the new version of the component, such that redundant delta operations can be removed. The new application order constraint of the delta DElectronicStability-Control is (**DAntiLockBrakingSystem** && **!DFourWheelDrive**), since a reference to delta DTractionControl is no longer required. Afterwards, all product configurations containing the delta DTractionControl are adapted to only include the new version of delta DElectronicStabilityControl and delta DTractionControl is removed.

*Merge-With-Core-Refactoring.* The Merge-With-Core-Refactoring merges the core of a product line with the content of deltas to create a new core model.

**Situation:** After product line evolution, it can happen that the core itself is not a valid product anymore. All product variant configurations contain the same subset of deltas that transform the outdated core to a valid product variant.

**Mechanics:** The Merge-With-Core-Refactoring is carried out as follows:

1. If the core itself is no supported product variant, identify a delta sequence $D_1, ..., D_n$ that is directly applied to the core such that the intermediate products are also no supported product variants.
2. Apply the deltas $D_1, ..., D_n$ to the core to create a new core for the product line.
3. Adjust supported product variants by removing the deltas $D_1, ..., D_n$.
4. Adjust application conditions of remaining deltas by removing the deltas $D_1, ..., D_n$.
5. Remove the deltas $D_1, ..., D_n$ that are now integrated into the core from the product line.

**Effect:** After applying this refactoring, the core is valid product again. By reducing the amount of available deltas, comprehensibility of the product line has been increased while decreasing overall complexity.

```
component BrakingSystem {
  autoconnect port;

  port
    in BrakeCommand brake,
    out BrakePressure wheelpressure1,
    out BrakePressure wheelpressure2,
    out BrakePressure wheelpressure3,
    out BrakePressure wheelpressure4,
    in WheelSensor wheelspeed1,
    in WheelSensor wheelspeed2,
    in WheelSensor wheelspeed3,
    in WheelSensor wheelspeed4;

  component ABS brakefunction;
}
```

**Listing 13.** Core containing delta `DAntiLockBrakingSystem`.

**Example:** In our case example (see Fig. 10), the core does not represent a supported product variant any more. Delta `DAntiLockBrakingSystem` has to be applied to the core before we obtain the product variant `CarWithABS`. Hence, the delta `DAntiLockBrakingSystem` can be integrated into the core. Fig. 14 shows the structure of the product line after applying the Merge-With-Core-Refactoring and the previous Compose-Deltas-Refactoring. The new core architecture is shown in Lst. 13.

*Merge-With-Core-Refactoring With Inverse Deltas.* In some cases, it can be useful to integrate a sequence of deltas into the core, although there is a product variant that is represented by the existing core.

**Situation:** A reason for this scenario may be that in the future the new core will become the basis for product development, but the old core should still be maintained for a transitional period of time. After this time, it should be possible to easily remove the old core from the product line. The respective sequence of deltas can already be integrated into the new core, if the old core is reconstructable as long as necessary.

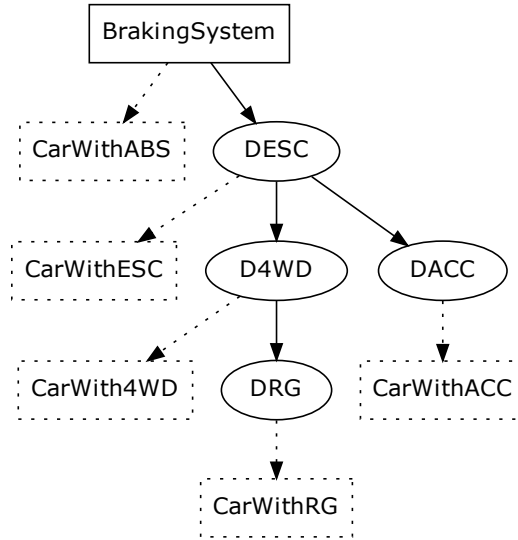

**Fig. 14.** Deltas and configurations after two refactorings

**Mechanics:** This transformation of the product line can be achieved using inverse deltas. An *inverse delta* [15] is a delta which reverts modifications carried out by another delta. An inverse delta of some existing delta is created by changing add operations to remove operations and vice versa. Modification operations have to be handled separately depending on the structure they alter. In [15], we show that for every delta in $\Delta$-MontiArc a corresponding inverse delta exists.

A Merge-With-Core-Refactoring with Inverse Deltas is performed as follows (where the first 4 steps perform a Merge-With-Core-Refactoring):

1. Identify a delta sequence $D_1, ..., D_n$ which should be integrated into the core. The core represents an existing product, while there are no intermediate products generated by the delta sequence that correspond to supported product variants.
2. Apply the deltas $D_1, ..., D_n$ to the core to create a new core for the product line.
3. Update the remaining product configurations and the application order constraints of the remaining deltas by removing any references to the deltas $D_1, ..., D_n$.

4. Remove the deltas $D_1, ..., D_n$ from the product line.
5. Create an inverse delta for the sequence of deltas $D_1, ..., D_n$ by inverting the delta operations of the delta that is obtained by composing the sequence of deltas $D_1, ..., D_n$ (as described in the Compose-Deltas-Refactoring). The application order constraint of the inverse delta is the negation of all other deltas such that the inverse delta is always applied first in any product configuration. This delta transforms the new core to the old core. Although the application order constraint for the delta is not needed for this particular scenario, it is useful for further evolution steps.
6. Add a product configuration for obtaining the old core which only contains the inverse delta.

**Effect:** The refactoring merges a set of mostly used deltas with the core. For products that do not contain these deltas, the old core may be reconstructed by applying the created inverse delta. It is usefull, if the refactored deltas are part of the majority of product variants and the other products will be removed from the product line anytime soon. This way, development of new product variants is eased, as they may be build up on a richer core model.

```
delta DInverse after !DAdaptiveCruiseControl
    && !DFourWheelDrive && !DReductionGear {
  modify component BrakingSystem {
    remove port accel;
    remove port lateralaccel;
    remove component stabilizer;

    disconnect brakefunction.wheelpressure1 ->
        stabilizer.fromabs1;
    disconnect brakefunction.wheelpressure2 ->
        stabilizer.fromabs2;
    disconnect brakefunction.wheelpressure3 ->
        stabilizer.fromabs3;
    disconnect brakefunction.wheelpressure4 ->
        stabilizer.fromabs4;
  }
}
```
**Listing 15.** Inverse delta for delta `DElectronicStabilityControl`.

**Example:** Assume that the variant `CarWithESC` should become the new core since every new car in the near future should be equipped with an electronic stability control. Hence, delta `DElectronicStabilityControl` shown in Lst. 12 can be integrated into the core. The new core shown in Lst. 13 is now serving as basis for all product variant generation. All product variants that previously used delta `DElectronicStabiliyControl` are adjusted as well as the application order constraints of the remaining deltas. The delta `DElectronicStabiliyControl` is removed from the product line. However, variant `CarWithABS` should still be supported for a transitional period of time. Hence, an inverse delta is required that reverts the modifications of delta `DElectronicStabilityControl`. This new delta `DInverse` is shown in Lst. 15. It is added to the product line, and a new product configuration for the product variant `CarWithABS` is added that applies the inverse delta to the new core. The resulting product line is depicted in Fig. 16.

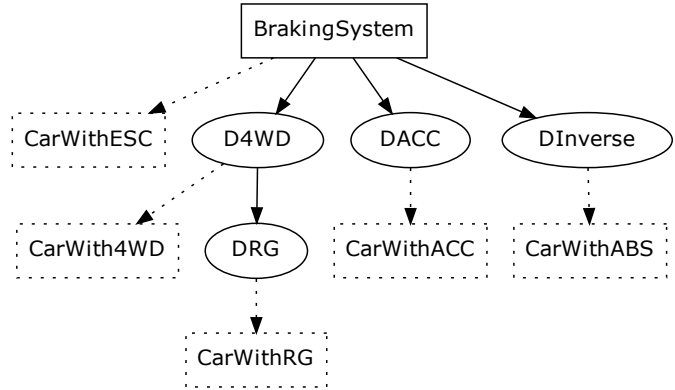

**Fig. 16.** Deltas and configurations after refactoring with inverse deltas

The concept of inverse deltas is very flexible. Hence, it is possible to always include features into the core architecture and add inverse deltas to the product line to remove these features in order to generate specific product variants not containing these features. This is particularly advantageous if the core architecture is one of the main products of the product line since it can be thoroughly validated and verified using standard single application engineering techniques.

## 5 Comparison to Annotative Variability Modeling

The predominantly used approach in industrial applications for modeling architectural variability is annotative variability modeling [39]. Our experience shows that annotative variability modeling is the easiest way to add variability information to an existing software product. However, a subsequent change to another variability modeling method is mostly not realized since this is often very time consuming. Annotative variability modeling is based on a 150%-model capturing all possible variability and annotating specific elements with the variant(s) in which they are included. Elements of the core architecture have no annotations. In order to derive a particular variant, all elements annotated with only different variants are removed.

In order to compare $\Delta$-MontiArc and its capabilities to capture product line evolution with annotative variability modeling, we realized all scenarios in

Sec. 3 and Sec. 4 with $\Delta$-MontiArc and also with annotative variability modeling. We decided to compare $\Delta$-MontiArc with an annotative modeling approach for MontiArc, since annotative variability modeling is the main variability modeling approach used current industrial practice. An annotative MontiArc dialect offers a good comparability to $\Delta$-MontiArc, as both langauges are based on the same syntax and exclusively differ in its variability modeling technique. For our comparison, we do not consider compositional variability modeling approaches, such based on aspect-oriented implementation techniques [1, 10], since these approaches do not natively support extractive product line development and the removal of modeling elements. The ability to explicitly represent removals is, however, essential for the direct representation of product line evolution without considering additional changes in the model structure, e.g., by refactorings before the evolution is carried out. Tool support for both modeling approaches, annotative and delta-based, is provided by the MontiCore framework for developing domain-specific languages [13] by extending the existing implementation of MontiArc [18].

*Annotative Variability Modeling in MontiArc* For realizing annotative variability modeling in MontiArc, each architectural element is annotated by a stereotype denoting the variant(s) in which it is included. Variable parts of an architecture are ports, subcomponents, and connectors. Hence, these elements may be annotated to assign them to variants. The excerpt of an annotated MontiArc model in Lst. 17 shows an example of these stereotype annotations for architectural elements. Line 2 contains an incoming port without any annotation indicating that this element is part of the core architecture. The incoming port in l. 4 is only needed for bikes. The corresponding annotation in l. 3 states that this incoming port is only used in the variant `BikeWithABS`.

```
port
   in BrakeCommand brake,
   <<variant = "BikeWithABS">>
   in BrakeCommand brakerear;
```

Listing 17. Example of model element annotation.

In annotative variability modeling, *adding a product variant* to the product line means to add new architectural elements to the 150%-model and to annotate these and already existing architectural elements with the newly added variant. This can require to change the 150%-model in several places and might become fairly complex not to miss necessary additions. In delta modeling, simply new deltas and product configurations can be added that locally encapsulate the necessary modifications.

*Removing variants* in the annotative approach amounts to removing the respective variant annotations and also the architectural elements that are no longer required by any other variant. Here, again changes all over the variability model may be necessary. Also, it has to be taken care that architectural elements belonging to the core without annotations are not accidentally removed and that architectural elements of removed variants are not silently added to the core. In delta modeling, variants are removed locally by changing the respective product configurations and deleting redundant deltas.

The *modification of existing variants* in the annotative approach can have an impact on several architectural elements. New elements are added and annotated with the specific variant and redundant elements are removed. This is particularly difficult, since variants which are not affected by the modification should not be changed. In delta modeling, only the content of specific deltas has to be changed while the application order constraints, the other deltas and the product configurations remain unchanged.

The *refactorings* presented in Sec. 4 are specifically tailored to $\Delta$-MontiArc. In particular, the Compose-Deltas-Refactoring and the Merge-With-Core-Refactoring with Inverse Deltas can only be applied in delta modeling. However, also in annotative variability modeling, it is possible to move certain variants to the core. This requires to determine all architectural elements which should belong to the core in the future. In a subsequent step, all annotations referring to these variants can be removed. This again might be a fairly complex and error-prone task, since it requires modifications in all parts of the variability model where architectural elements belonging to the considered variants occur. In delta modeling, only the deltas which should be included in the core have to be integrated, and application conditions of other deltas and specific product configurations can be changed locally.

*Comparison with $\Delta$-MontiArc* The modeling of the product line evolution scenarios in the annotative approach is very time consuming and error-prone since changes to product variants or the core architecture potentially require changes in all parts of the product line model. For every architectural modeling element, it has to be decided in which specific variants it appears. In delta modeling, changes are encapsulated modularly in deltas and can be performed locally. While in the annotative variability modeling approach, variability modeling is mixed with modeling of the functional architecture, in delta modeling, variability is a first-class entity. Deltas only focus on the representation of variability and are, thus, easier to comprehend and to evolve.

In order to quantitatively compare $\Delta$-MontiArc with annotative modeling, we consider all implementations which are modeled in out case study. In total, we look at seven different product line scenarios: the base scenario is the initial product line depicted in Fig. 5; the first scenario is the product line after adding the product variant `CarWithRG`; the second scenario corresponds to the product line after removing variants which is depicted in Fig. 10 (p. 10); the third scenario is the product line after modification of variant `CarWithACC`. Scenarios 4 to 6 correspond to the three different refactoring strategies: the fourth scenario is the product line after the Compose-Deltas-Refactoring; the fifth scenario is the product line after also applying the Merge-With-Core-Refactoring as it is depicted in Fig. 14 (p. 14); the sixth scenario is the product line after the Merge-With-Core-Refactoring With Inverse Deltas depicted in Fig. 16 (p. 16).

|  | base | | scenario 1 | | scenario 2 | | scenario 3 | |
|---|---|---|---|---|---|---|---|---|
|  | $\Delta$ | 150% | $\Delta$ | 150% | $\Delta$ | 150% | $\Delta$ | 150% |
| #components | 6 | | 7 | | 6 | | 5 | |
| #ports | 67 | | 75 | | 56 | | 55 | |
| #connections | 6 | | 10 | | 10 | | 10 | |
| #variants | 8 | | 9 | | 5 | | 5 | |
| #chars | 4209 | 4156 | 5048 | 5111 | 3887 | 4056 | 3803 | 3956 |
| #varchars | 2437 | 1591 | 2954 | 1916 | 2238 | 1472 | 2264 | 1456 |
| rel. variant inf. | 57% | 38% | 58% | 37% | 57% | 36% | 59% | 36% |
| #files | 20 | 6 | 23 | 7 | 17 | 6 | 16 | 5 |
| #maxchars | 474 | 2087 | 474 | 2660 | 438 | 2284 | 438 | 2334 |
| avg. chars p. file | 210 | 692 | 219 | 730 | 228 | 676 | 237 | 791 |

**Table 18.** Quantitative comparisson of $\Delta$-MontiArc and annotative modeling for temporal variability

|  | scenario 4 | | scenario 5 | | scenario 6 | |
|---|---|---|---|---|---|---|
|  | $\Delta$ | 150% | $\Delta$ | 150% | $\Delta$ | 150% |
| #components | 5 | | 5 | | 5 | |
| #ports | 55 | | 55 | | 55 | |
| #connections | 10 | | 10 | | 10 | |
| #variants | 5 | | 5 | | 5 | |
| #chars | 3586 | 3956 | 3273 | 3219 | 3448 | 3219 |
| #varchars | 2047 | 1456 | 1649 | 719 | 1514 | 719 |
| rel. variant inf. | 57% | 36% | 50% | 22% | 43% | 22% |
| #files | 15 | 5 | 14 | 5 | 14 | 5 |
| #maxchars | 448 | 2334 | 438 | 1865 | 645 | 1865 |
| avg. chars p. file | 239 | 791 | 233 | 643 | 246 | 643 |

**Table 19.** Quantitative comparisson of $\Delta$-MontiArc and annotative modeling for refactoring scenarios

Tab. 18 and 19 show the results of our evaluation. For the overall sizes of the product lines in the different scenarios, we counted the total number of components (#components), ports (#ports), explicit connections (#connections), and supported product variants (#variants). Implicit connections created by the `autoconnect` statements are not counted. In the table, we see that all examples are mid-sized with 5 to 7 components and 5 to 9 supported product variants.

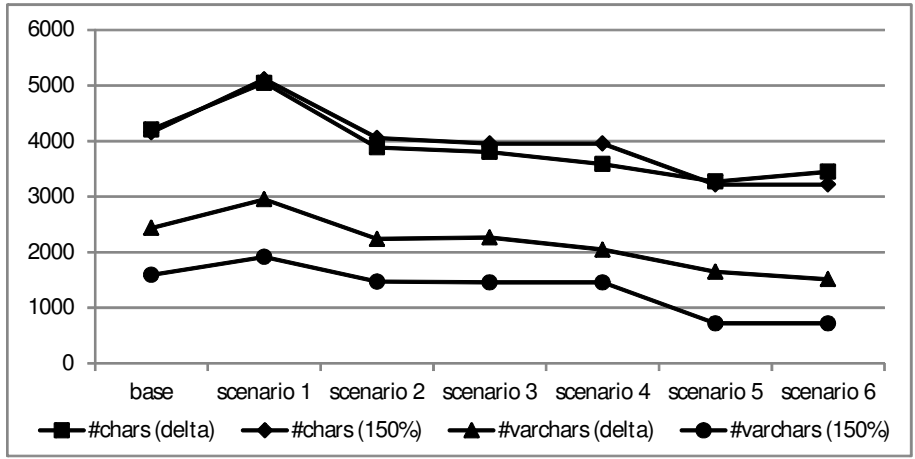

**Fig. 20.** Number of characters for model and variability representation.

For quantitatively comparing the way variability is encoded in delta modeling and annotative variability modeling, we measured the overall sizes of the models, the total amount of variability information required to express all product variants, and the relative amount of variability information compared to the information necessary for encoding of functionality. We computed the overall model sizes by counting each visible character (except for comments) in the product line model (#chars). Since MontiArc allows many different formatting styles, visible characters give a more accurate measure of the model size than lines of code. Also for the variability information, we counted the number of characters (#varchars) used for specifying deltas in $\Delta$-MontiArc-models and the characters used for annotations in the annotative variability model. To compare the ratio between variability and functional parts of the models, the relative amount of variant information is calculated (rel. variant inf.) by dividing the number of characters used for encoding variability by the total number of characters used in the overall model. These metrics are suitable, as both languages use the same syntax and exclusively differ in modeling variability.

Fig. 20 visualizes the overall number of characters used for representing the product line architectures in the different scenarios and the number of characters used to specify variability, both for delta modeling and annotative variability modeling. Roughly, we can say that the sizes of both models are the same for both variability modeling approaches. Adding product variants in the first scenario increases the size of the model and also the amount of variability information in both approaches. Removing variants decreases the size of the model and the variability information. Modification of an existing product variant only changes the size of the model and the amount of variability information slightly. In the fourth scenario, which is the first refactoring scenario, we see the advantage for delta modeling if deltas are combined. The overall size of the model after refactoring is lower than for the annotative variability model. Since the Compose-

Deltas-Refactoring and the Merge-With-Core-Refactoring with Inverse Deltas are not applicable for annotative variability modeling, the figures do not change from the third to the fourth scenario and from the fifth to the sixth scenario. The Merge-With-Core-Refactoring in the fifth scenario reduces the size of the model and also the amount of variability information for both approaches such that the model size is almost equal again. In scenario 6, after the Merge-With-Core-Refactoring with Inverse Deltas the size of the delta model is larger since the inverse delta is added to the product line model.

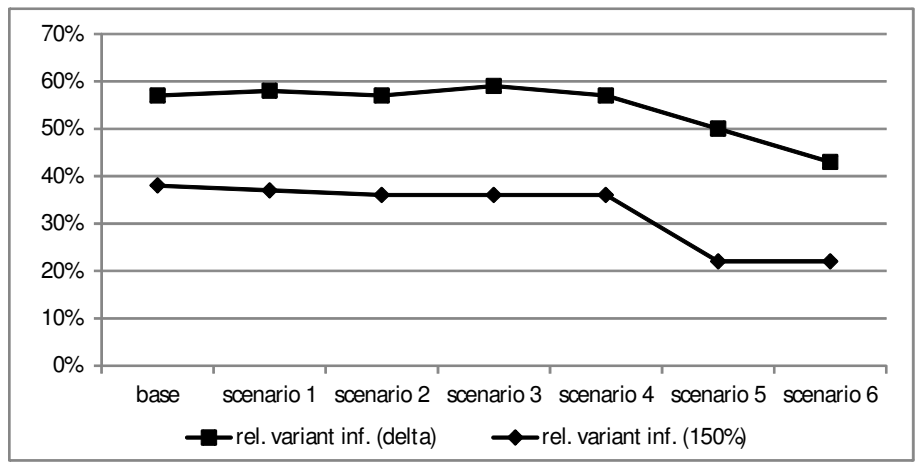

**Fig. 21.** Relative amount of variability information.

The ratios of variability information compared to the overall model size stay roughly the same for both approaches independent of the evolution scenario for the base and the first four scenarios as it is visualized in Fig. 21. Only, when variants are merged into the core the ratio drops. The reason for this is that by merging deltas with the core or removing annotations variability information is removed while the overall size of the model stays the same. In general, the ratio of variability information is higher for delta modeling than for annotative variability modeling. From the figures, delta modeling seems to be very similar to annotative variability modeling when comparing the mere model size and its changes for the different evolution scenarios. However, when looking at the ratio of variability information, we can see that in annotative variability modeling roughly one third of all characters are used for expressing variability by annotations. If we now consider that in the annotative variability modeling approach we only have one 150%-model of all possible variability and that variability information is spread all over the model, having a third of all characters used for annotations renders the model very complex and difficult to comprehend and evolve. In contrast, in delta modeling, variability information is encapsulated in deltas which can be evolved locally.

To further analyze understandability and maintainability, we measured the average sizes of the files which make up the product line models for the single scenarios (see Tab. 18 and 19). A product line model is distributed over several files (#files) where each file defines a component, a delta or parts of it. In general, large files are harder to understand, to change and to maintain. Hence, we measured the maximum (#maxchars) and average characters per file (avg. chars p. file). We can see that in delta modeling the number of files is generally higher which results from the fact that each delta and the contained components are stored in separate files while in annotative variability modeling only each component has a separate file. Overall, the size of a file in annotative variability modeling is three times larger than in delta modeling. Thus, the evaluation of the scenarios yields that delta modeling improves the readability and understandability of product line architectures and eases their evolution and maintenance by modularizing variability in small encapsulated entities.

## 6 Related Work

Most modeling approaches for architectural variability only consider one dimension of variability. For spatial variability, we can distinguish three main approaches: annotative, compositional and transformational variability modeling. Annotative approaches use variant annotations, e.g., UML stereotypes in UML models [40, 12] or presence conditions [7], to define which architectural elements belong to specific product variants. In the orthogonal variability model (OVM) [32], a separate variability representation with links to the architecture model replaces direct annotations. While annotative variability modeling allows fine-grained modifications, it relies on a monolithic product line representation.

Compositional approaches for modeling architectural variability [39] capture architectural variation by selecting specific component variants. In [10], Plastic partial components [31] model component variability by extending partially defined components with variation points and associated variants. Hierarchical variability modeling for software product lines [17] aims at combining component variability with the component hierarchy to foster component-based development of diverse systems during architectural design. Compositional variability modeling allows a modular description of variability, but limits the impact of changes to the applied composition technique.

Transformational approaches, such as delta modeling [6], represent variability by transformation of a base architectural model. In the common variability language (CVL) [19], elements of the base model are substituted according to a set of pre-defined rules. In [21], graph transformation rules capture the variability of a single kernel model comprising the commonalities of all systems. In [20], architectural variability is represented by change sets containing additions, removals or modifications of components and component connections that are applied to a base line architecture. All these approaches are only consider variability in space as the previous versions of $\Delta$-MontiArc [16, 14].

Temporal variability is usually specified with two mechanisms [29]: logical assertions or graph transformations. In the assertion-based approaches, e.g., [30, 38], a transformation is characterized by a pre-condition defining when a transformation can be applied and a post-conditions specifying the properties that are ensured by the transformation. In graph transformation-based approaches, the product variants are represented by graphs. System evolution is specified by a graph transformation rule, see e.g. [28]. These approaches, however, represent temporal variability on a meta-level.

In order to be able to reason about architectural evolution, it has to be captured as first-class entity [27]. One approach towards this goal [24] defines new components by explicitly expressing the differences to the old component by adding, deleting, renaming or replacing elements. This is very similar to delta modeling where a delta encapsulates the differences from one product variant to the other. Aspect-oriented composition is also applied to model software architecture evolution [3] expressing variability by weaving selected aspects into a core architecture. However, these approaches only consider architectural evolution.

Refactorings of feature-oriented programming (FOP) product lines are presented in [37]. These refactorings that move fields or types between feature modules are mostly based on classical code refactorings like, e.g., pulling up fields or methods to parent types. In alignment with our approach, the authors suggest refactorings of a product line to be variant-preserving. Hence, such a refactoring only changes the structure of the product line, but not contained variants. However, the presented approach aims at the implementation of a software product line and not at its architecture.

In contrast to the above approaches, product line evolution [9] considers the combination of variability in space and variability in time. Extractive product line engineering [22] develops a product line from a set of legacy applications; the proactive approach aims at evolving an initial product line if new user requirements arise. In the PuLSE product line engineering methodology [4], product line evolution is defined as designated development phase. However, these approaches only focus on terminological issues and development processes. There is some work on feature model evolution [5] and evolution of feature-oriented [2] or aspect-oriented software product line implementations [1, 10]. However, evolution in feature-oriented modeling and programming approaches is treated with different linguistic means than spatial variability, mostly due to the fact that features cannot remove model or program entities which is essential for capturing unexpected changes caused by evolution. Hence, a uniform modeling framework for architectural variability in space and in time is missing despite techniques, such as aspect-oriented composition and model transformations, that can factually express both dimensions of variability. $\Delta$-MontiArc fills this gap by representing variability in space and in time as a first-class entity with the same linguistic means.

## 7 Conclusion

We have proposed $\Delta$-MontiArc, an ADL with native support for variability based on delta modeling. $\Delta$-MontiArc allows expressing architectural variability in space and in time in modular and easily maintainable manner as we demonstrated by a quantitative and qualitative comparison with annotative variability modeling. We presented exemplary refactorings that help cleaning up a degenerated product line. Variability by using delta modeling can also be applied to other modeling languages and is not restricted to modeling software architectures. Behavioral variability within the architectural descriptions can be realized by using deltas on state machines [23] or Java source code [34, 36].

For future work, we aim at defining further refactorings that merge deltas with identical modification operations but different application order constraints and vice versa. Scalability and applicability has to be checked based on a more complex industrial-scale examples. We also plan to extend the conceptual ideas of $\Delta$-MontiArc into a seamless software engineering process for software product lines that allows dealing with variability in space and in time by the same techniques.